\documentclass[twocolumn,prb,color,psfig,superscriptaddress]{revtex4}
\usepackage{graphicx}
\usepackage{epsfig}
\begin{document}

\title{Topological phase separation  in 2D
hard-core Bose-Hubbard system away from half-filling}
\author{A.S. Moskvin}
\author{I.G. Bostrem}
\author{A.S. Ovchinnikov}
\affiliation{Ural State University, 620083, Ekaterinburg,  Russia}

\date{\today}
\begin{abstract}
We  suppose that the doping of the 2D
hard-core boson system away from half-filling may result in the formation of
multi-center topological defect such as charge order (CO) bubble domain(s) with
Bose superfluid (BS) and extra bosons both localized in domain wall(s), or a
{\it topological} CO+BS {\it phase
separation}, rather than an uniform mixed CO+BS supersolid phase. Starting from
the classical model we predict the properties of the respective quantum system.
The long-wavelength behavior of the  system is believed to remind that of
granular superconductors, CDW materials,  Wigner crystals, and multi-skyrmion
system akin in a quantum Hall ferromagnetic state of a 2D electron gas.
\end{abstract}

\maketitle


One of the fundamental hot debated  problems in bosonic physics concerns the
evolution of the charge ordered (CO) ground state of 2D hard-core Bose-Hubbard
(BH) model (hc-BH)  with a doping away from half-filling.
Numerous model studies steadily confirmed the emergence of "supersolid" phases
with simultaneous diagonal and off-diagonal long range order while Penrose and
Onsager \cite{Penrose} were the first showing as early as 1956 that supersolid
phases do not occur.

The most recent quantum Monte-Carlo  (QMC) simulations
\cite{Batrouni,Hebert,Schmid}
found two significant features of the 2D BH model with a
 screened Coulomb repulsion: the absence of supersolid
phase  at half-filling, and a  growing tendency to phase separation (CO+BS)
upon doping away from half-filling.  Moreover, Batrouni and Scalettar
\cite{Batrouni} studied quantum phase transitions in the ground state
of the 2D hc-BH Hamiltonian and have shown  numerically that,
contrary to the generally held belief, the most commonly discussed
"checkerboard" supersolid is thermodynamically unstable and phase separates
into solid and superfluid phases. The physics of the CO+BS phase separation in
BH model is associated with a rapid increase of the energy of a
homogeneous CO state with doping away from half-filling due to a large
"pseudo-spin-flip" energy cost.
Hence, it appears to be
energetically more favorable to "extract" extra bosons (holes) from the CO
state and arrange them into finite clusters with a relatively small number of
particles. Such a droplet scenario is believed  to minimize the long-range
Coulomb repulsion.\cite{Cagan}
  However, immediately there arise several questions. Whether a simple
mean-field approximation  (MFA)
and classical continuum model can predict such a behavior? What is the detailed
structure of the CO+BS phase separated state? What are the main factors
governing the essential low-energy and long-wavelength physics? Is it possible
to make use of
simple toy models?

In the Letter we
present a topological scenario of CO+BS phase separation in 2D hc-BH model with
inter-site repulsion. The extra
bosons/holes doped to a checkerboard CO phase of 2D boson system are believed
to be confined in the ring-shaped domain wall of the skyrmion-like topological
defect which looks like a bubble domain in an easy-axis antiferromagnet. This
allows us to
propose the mechanism of 2D {\it topological} CO+BS {\it phase separation} when
the
doping of the bare checkerboard CO phase results in the formation of a
multi-center topological defect, which simplified pseudo-spin pattern looks
like  a system of bubble CO domains with Bose superfluid confined in  charged
ring-shaped domain walls.


The Hamiltonian of hard-core Bose gas on a lattice can be written in a standard
form as follows:
\begin{equation}
\smallskip
H_{BG}=-\sum\limits_{i>j}t_{ij}{\hat
P}(B_{i}^{\dagger}B_{j}+B_{j}^{\dagger}B_{i}){\hat P}
+\sum\limits_{i>j}V_{ij}N_{i}N_{j}-\mu \sum\limits_{i}N_{i},  \label{Bip}
\end{equation}
where ${\hat P}$ is the projection operator which removes double occupancy of
any site,  $N_{i}=B_{i}^{\dagger}B_{i}$, $\mu $  the chemical potential
determined from the condition of fixed full number of bosons $N_{l}=
\sum\limits_{i=1}^{N}\langle N_{i}\rangle $ or concentration $\;n=N_{l}/N\in
[0,1]$. The $t_{ij}$ denotes an effective transfer integral,  $V_{ij}$ is an
intersite interaction between the bosons. \smallskip Here $B^{\dagger}(B)$ are
the Pauli creation (annihilation) operators which are Bose-like commuting for
different sites $[B_{i},B_{j}^{\dagger}]=0,$ for $i\neq j,$ while  for the same
site $B_{i}^{2}=(B_{i}^{\dagger})^{2}=0$, $[B_{i},B_{i}^{\dagger}]=1-2N_i$, $N_i
= B_{i}^{\dagger}B_{i}$; $N$ is a full number of sites.

\smallskip The model of hard-core Bose-gas with an intersite repulsion is
equivalent to a system of spins $s=1/2$  exposed to an external magnetic field
in the $z$%
-direction
\begin{equation}
H_{BG}=\sum_{i>j}J^{xy}_{ij}(s_{i}^{+}s_{j}^{-}+s_{j}^{+}s_{i}^{-})+\sum\limits_
{i>j}
J^{z}_{ij}s_{i}^{z}s_{j}^{z}-h \sum\limits_{i}s_{i}^{z}, \label{spinBG}
\end{equation}
where $J^{xy}_{ij}=2t_{ij}$, $J^{z}_{ij}=V_{ij}$, $h =\mu -\sum\limits_{j
(j\neq i)}V_{ij}$, $s^{-}= \frac{1}{\sqrt{2}}B_ , s^{+}=-\frac{1}{\sqrt{2}}
 B^{\dagger}, s^{z}=-\frac{1}{2}+B_{i}^{\dagger}B_{i}$,
$s^{\pm}=\mp \frac{1}{\sqrt{2}}(s^x \pm \imath s^y)$.

In frames of of a conventional two-sublattice MFA approach one may introduce two
vectors, the ferromagnetic and antiferromagnetic ones:
$$
{\bf m}=\frac{1}{2s}(\langle{\bf s}_1 \rangle +\langle{\bf s}_2 \rangle);\,
{\bf l}=\frac{1}{2s}(\langle{\bf s}_1 \rangle -\langle{\bf s}_2
\rangle);\,\,{\bf m}^2 +{\bf l}^2 =1\, .
$$
where ${\bf m}\cdot {\bf l}=0$. For the plane where these vectors lie one may
introduce two-parametric angular description: $m_x = \sin \alpha \cos\beta ,m_z
= -\sin \alpha \sin\beta , l_x = \cos \alpha \sin\beta , l_z = \cos \alpha
\cos\beta $.
The hard-core boson system in a two-sublattice approximation is described by
two diagonal order parameters $l_z ,m_z$ and two complex off-diagonal
 order parameters $m_{\pm}=\mp \frac{1}{\sqrt{2}}(m_x \pm \imath m_y)$ and
$l_{\pm}=\mp \frac{1}{\sqrt{2}}(l_x \pm \imath l_y)$. The complex superfluid
order parameter $\Psi ({\bf r})=|\Psi ({\bf r})|\exp-\imath\varphi $ is
determined by the in-plane components of ferromagnetic vector: $ \Psi ({\bf
r})=\frac{1}{2}\langle (\hat B_1 +\hat B_2 )\rangle
=sm_{-}=sm_{\perp}\exp-\imath\varphi $, $m_{\perp}$ being the length of the
in-plane component of ferromagnetic vector. So, for a local condensate density
we get $n_s = s^2 m_{\perp}^2$.  It is of interest to note that in
fact all the conventional uniform $T=0$ states with nonzero $\Psi ({\bf r})$
imply simultaneous long-range order both for modulus $|\Psi ({\bf r})|$ and
phase $\varphi$. The in-plane components of antiferromagnetic vector $l_{\pm}$
describe a staggered off-diagonal order. It is worth noting that by default one
usually considers the negative sign of the transfer integral $t_{ij}$, that
implies the ferromagnetic in-plane pseudospin  ordering.

  The model exhibits many fascinating quantum phases and phase
transitions. Early investigations predict at $T=0$ charge order (CO), Bose
superfluid (BS) and mixed (CO+BS) supersolid uniform phases with an Ising-type
melting transition (CO-NO) and Kosterlitz-Thouless-type (BS-NO) phase
transitions to a non-ordered normal fluid (NO).\cite{MFA}
The detailed mean-field and spin-wave analysis of the uniform phases of 2D hc-BH
model is given by Pich and Frey.\cite{Pich}

  MFA yields for the
conventional uniform supersolid  phase \cite{Kubo}
$$
\sin^2 \beta = m_z \frac{\sqrt{V-2t}}{\sqrt{V+2t}}; \quad \sin^2 \alpha = m_z
\frac{\sqrt{V+2t}}{\sqrt{V-2t}}
$$
with a constant chemical potential $\mu = 4s\sqrt{(V^2 -4t^2)}$. It
should be noted that the supersolid phase appears to be unstable
with regard to small perturbations in the Hamiltonian. The
mean-field energy per site of the uniform supersolid phase is
written as follows:\cite{Pich}
$$
E_{SS}=E_{CO}+s\mu m_z = E_{CO}+\mu (n_B -\frac{1}{2}),
$$
where $E_{CO}=-2Vs^2$. The cost of doping both for CO and CO+BS phase appears
to be rather high as compared with an easy-plane BS phase at half-filling where
the chemical potential turns into zero.\cite{Bernardet}

Magnetic analogy allows us to make unambiguous predictions as
regards the doping of BH system away from half-filling. Indeed,
the boson/hole doping of checkerboard CO phase corresponds to the magnetization
of an antiferromagnet in $z$-direction. In the uniform easy-axis $l_z$-phase of
anisotropic antiferromagnet the local spin-flip energy cost is rather big. In
other words, the energy cost for boson/hole doping into  checkerboard CO phase
appears to be big due to a large contribution of boson-boson repulsion.

However, the  magnetization of the  anisotropic antiferromagnet in an easy axis
direction may proceed as a first order phase transition with a ``topological
phase separation'' due to the existence of antiphase domains.\cite{AFM-domain}
The antiphase domain walls provide the natural nucleation  centers for a
spin-flop phase having enhanced magnetic susceptibility as compared with small
if any longitudinal susceptibility thus  providing the advantage of the field
energy. Namely domain walls  would specify the inhomogeneous magnetization
pattern for such an  anisotropic  easy-axis  antiferromagnet in relatively weak
external magnetic field. As concerns the domain type in quasi-two-dimensional
antiferromagnet one should emphasize the specific role played by the
cylindrical, or bubble domains which have finite energy and size. These
topological solitons have the vortex-like in-plane spin structure and often
named "skyrmions". The classical, or Belavin-Polyakov (BP) skyrmion
\cite{Belavin} describes the
solutions of a non-linear $\sigma$-model with a classical isotropic 2D
Heisenberg
Hamiltonian and represents one of the generic toy model spin textures.
It is of primary importance  to note that skyrmion-like bubble domains in
easy-axis layered antiferromagnets  were actually observed  in the experiments
performed by Waldner,\cite{Waldner} that were  supported later  by different
authors (see, e.g. Refs.\onlinecite{Kochelaev,Carsten}). Although  some
questions were not completely clarified and remain open until now,
\cite{Kamppeter,Sheka} the classical and quantum  \cite{Istomin} skyrmion-like
topological defects are believed to be a genuine element of essential physics
both of ferro- and antiferromagnetic 2D easy-axis systems.

The magnetic analogy seems to be a little bit naive, however, it catches the
essential physics of doping the hc-BH system.
      As regards the checkerboard CO phase of such a  system, namely a  finite
energy  skyrmion-like  bubble domain \cite{AFM}     seems to be the  most
preferable candidate for the  domain with antiphase domain wall providing the
natural reservoir for extra bosons.
The classical description of nonlinear excitations in hc-BH model  includes the
skyrmionic solution given $V=2t$. \cite{AFM}
The skyrmion spin texture
consists of a vortex-like
arrangement of the in-plane components of ferromagnetic ${\bf m}$ vector  with
the $l_z$-component reversed
in the centre of the skyrmion and gradually increasing to match the homogeneous
background at infinity. The simplest spin distribution within classical skyrmion
 is given as follows
$$
m_{x}=m_{\perp}\cos (\varphi + \varphi _0) ;\,
m_{y}=m_{\perp}\sin (\varphi + \varphi _0);
$$
\begin{equation}
m_{\perp}=\frac{2r\lambda }{r^{2}+\lambda ^{2}} ;
l_z=\frac{r^{2}-\lambda^{2}}{r^{2}+\lambda^{2}},
\label{sk}
\end{equation}
where $\varphi _0$ is a global phase ($U(1)$ order parameter), $\lambda$
skyrmion radius. The skyrmion spin texture  describes the
coexistence and competition of the charge order parameter $l_z$ and BS order
parameter $m_{\perp}$ that reflects a complex spatial interplay of potential and
kinetic  energies.
The skyrmion looks like a bubble domain in an easy-axis magnet.
Skyrmions are characterized by the magnitude and sign of its topological
charge, by its size (radius), and by the global orientation of the spin. The
scale invariance of classical BP skyrmionic solution reflects in that its energy
 does not
depend on  radius, and global phase.
An interesting example of topological inhomogeneity is provided by a
multi-center
BP skyrmion \cite{Belavin} which energy does not depend on the
position of the centers. The latter are believed to be addressed
as  an additional degree of freedom, or positional order
 parameter.

 It should be noted that  the domain wall in such a
bubble domain somehow created in the checkerboard CO phase of 2D hc-BH system
represents an effective ring-shaped reservoir both for Bose-superfluid and extra
boson/hole.
Indeed, the magnetic response of a fixed radius skyrmion to a weak longitudinal
external magnetic field can be  approximated as follows:
$$
m_{z}(r) \approx \chi (r)h\, ,
$$
where $\chi (r)=f(r)m_{\perp}(r)$ is a local magnetic susceptibility, and $f(r)$
has distinct maximum at the domain wall center $r=\lambda$ with  oscillating
behavior $f(r)\propto J_{1}(r\sqrt{2})\propto \sin(r\sqrt{2}+\pi /4)/\sqrt{r}$
for large $r$. The oscillating character of the magnetization distribution seems
to reflect a  skyrmion instability. Thus we conclude that the extra boson/hole
density described by the order parameter $m_z$ is to be localized in the
domain wall.
 This result is supported by  numerical calculations for the static
skyrmion-like excitations in a continuous 2D hc-BH model with a strong
boson-boson repulsion  that was considered by us earlier in
Ref.\onlinecite{AFM}.
 The Fig. 1  sketches the calculated radial  distribution of the order
  parameter   $m_z$   for  a single bubble domain in hc-BH system with injected
small boson concentration away from half-filling. It appears to be trapped
inside the domain wall. As expected, the soliton   energy depends quadratically
on the number $ \Delta n$ of bosons bound in
domain wall (see insert in Fig. 1), similarly to that of homogeneous BS
phase.\cite{Bernardet} In other words, $\partial E/\partial n =0$ and one might
say about a zero value of the effective boson/hole chemical potential for the CO
bubble domain configuration would it be a ground state.

In the continuous model the classical BP skyrmion is a topological excitation
and cannot
dissipate. However, the classical static skyrmion is unstable with regard to an
external field, anisotropic interactions both of easy-plane and easy-axis type.
 Small easy-axis anisotropy or external field are sufficient to shrink
skyrmion to a nanoscopic size when magnetic length $l_0$:
$$
l_0  =\left( \sqrt{\left(2V/t\right) ^{2}-\left( \mu /t\right) ^{2}}-4\right)
^{-\frac{1}{2}}
$$
is of the order of
several lattice parameters, and the continuous approximation fails to correctly
describe excitations.
Nonetheless, Abanov and Pokrovsky \cite{Abanov} have shown that the easy-axis
anisotropy together with fourth-order exchange term can stabilize skyrmion with
radius $R\propto \sqrt{l_0}.$
\begin{figure}[t]
\includegraphics[width=8.5cm,angle=0]{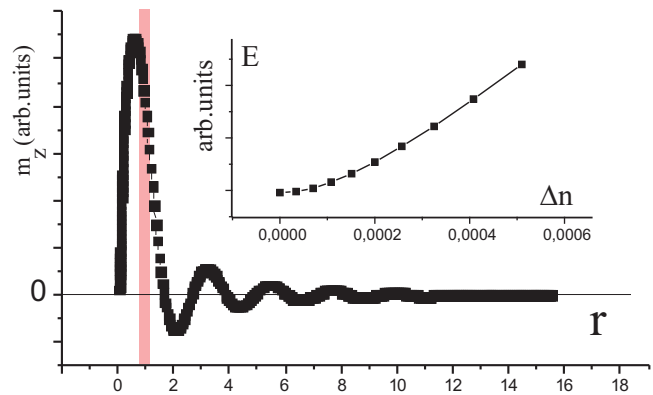}
\caption{The results of numerical calculations by shooting method
for a shrinked skyrmion given $v=2.1$, $t=1$: main panel - the
radial  distribution of the order  parameter   $m_z$ (the filled
region points to a center of a domain wall); the insert shows the
dependence of the soliton energy on the number $\Delta n$ of bosons
bound in a domain wall. } \label{fig1}
\end{figure}

A skyrmionic scenario in hc-BH model allows us to make several important
predictions.
 Away from half-filling  one may
anticipate the  nucleation of a topological defect in the unconventional form of
the multi-center skyrmion-like object with  ring-shaped Bose superfluid regions
positioned in an antiphase domain wall separating the CO core
 and CO outside of the single skyrmion. The specific spatial separation of BS
and CO order parameters that avoid each other reflects the competition of
kinetic  and potential energy. Such a {\it topological} (CO+BS) {\it phase
 separation} is believed to provide a minimization of the total energy as
compared with its uniform supersolid counterpart.
Thus, the  parent checkerboard CO phase may gradually loose its stability under
boson/hole doping, while a novel topological self-organized texture is believed
to become stable.
 The most probable possibility is that every domain wall accumulates single
boson, or boson hole. Then, the number of centers in a multi-center skyrmion
nucleated with doping  has to be equal to the number of bosons/holes.   In such
a case, we anticipate the near-linear dependence of the total BS volume fraction
on the doping.
 Generally speaking, one may assume scenario
 when the nucleation of a  multi-center skyrmion  occurs spontaneously
 with no doping. In such a case we should anticipate the existence of neutral
 multi-center skyrmion-like object with equal number of positively and
negatively charged single skyrmions. However, in  practice, namely the
boson/hole doping is likely to be a physically clear driving force
for a nucleation of  a single, or multi-center skyrmion-like self-organized
collective mode in the form of multi-center charged topological defect
 which may be (not strictly correctly) referred to as
multi-skyrmion system akin in a quantum Hall ferromagnetic state of a
two-dimensional electron gas.\cite{Green} In such a case, we may characterize an
individual
skyrmion by its position (i.e., the center of skyrmionic texture), its size
(i.e., the radius of domain wall), and the orientation of the in-plane
components of pseudo-spin (U(1) degree of freedom). An isolated skyrmion is
described by the inhomogeneous distribution of the CO parameter, or staggered
boson density $l_z$, order parameter $m_z$ characterizing the deviation from
the half-filling, and $m_{\perp}$ that corresponds only to the modulus of the
superfluid order parameter.

It seems likely that for a light doping any  doped particle (boson/holes)
results in a nucleation
of a new single-skyrmion state, hence its density changes gradually with
particle doping.
Therefore, as long as the separation between skyrmionic centers is sufficiently
large so that the inter-skyrmion interaction is negligible, the energy of the
system per particle remains almost constant. This means that the chemical
potential of boson or hole remains unchanged with doping and hence apparently
remains fixed.

 The multi-skyrmionic system in contrast with uniform ones can form the
structures
 with inhomogeneous long-range ordering of the modulus of the superfluid order
 parameter accompanied by the unordered global phases of single skyrmions.
Such a situation resembles in part that of granular superconductivity.

In the long-wavelength limit the off-diagonal ordering can be described by an
effective Hamiltonian in terms of  U(1) (phase) degree of freedom associated
with each skyrmion. Such a Hamiltonian
 contains a repulsive, long-range Coulomb part and a
short-range contribution related to the phase degree of freedom. The
latter term can be written out in the standard for the $XY$ model form of a
so-called Josephson coupling
\begin{equation}
H_J = -\sum_{\langle i,j\rangle}J_{ij}\cos(\varphi _{i}-\varphi _{j}),
\end{equation}
where $\varphi _{i},\varphi _{j}$ are global phases for skyrmions centered at
points $i,j$, respectively, $J_{ij}$ Josephson coupling parameter. Namely the
Josephson coupling gives rise to the long-range ordering of the phase of the
superfluid order parameter in a multi-center skyrmion. Such a Hamiltonian
represents a starting point for the analysis of disordered superconductors,
granular superconductivity, insulator-superconductor transition with $\langle
i,j\rangle$ array of superconducting islands with phases $\varphi _{i},\varphi
_{j}$. Calculating the phase-dependent part of skyrmion-skyrmion interaction
Timm {\it et al.}\cite{Timm} arrived at $negative$ sign of $J_{ij}$ that
 favors antiparallel alignment of the U(1) pseudospins. In other words,
 two skyrmions are assumed to form a peculiar Josephson  $\pi$ micro-junction.
There are a number of interesting implications that follow directly from this
result:\cite{Kivelson} the spontaneous breaking of time-reversal symmetry with
non-zero supercurrents and magnetic fluxes in the ground state, long-time tails
in the dynamics of the system, unconventional Aharonov-Bohm period $hc/4e$,
negative magnetoresistence.

To account for Coulomb interaction and allow for quantum corrections we should
introduce into effective Hamiltonian  the charging energy \cite{Kivelson}
$$
H_{ch}=-\frac{1}{2}q^2 \sum_{i,j}n_{i}(C^{-1})_{ij}n_{j}\, ,
$$
where $n_{i}$ is a boson number operator for bosons bound in $i$-th skyrmion; it
is quantum-mechanically conjugated to $\varphi$: $n_{i}=-i \partial /\partial
\varphi
_{i}$, $(C^{-1})_{ij}$ stands for  the capacitance matrix, $q$ for bosonic
charge.

 Such a system appears to reveal a tremendously rich quantum-critical structure.
In the absence of disorder, the
$T=0$ phase diagram of the multi-skyrmion system implies either triangular, or
square crystalline arrangements (Skyrmion crystal)
 with possible melting transition to a Skyrmion liquid.
 It should be noted that analogy with charged $2D$ Coulomb gas
implies the Wigner crystallization of  multi-center skyrmion with Wigner
crystal to Wigner liquid melting transition, respectively. Naturally, that the
additional degrees of freedom for skyrmion provide a richer physics of
Skyrmion
lattices. For a WC to be an insulator, disorder is required, which pins the WC
and also causes the crystalline order to have a finite correlation length.
Traditional approach to a Wigner crystallization implies the formation of a WC
for densities lower than a critical density, when the Coulomb energy dominates
over the kinetic energy. The effect of quantum fluctuations leads to a
(quantum) melting of the solid at high densities, or at a critical lattice
spacing. The critical properties of a two-dimensional lattice without any
internal degree of freedom are successfully described  applying the BKT theory
to dislocations and disclinations of the lattice, and proceeds in two steps. The
first implies the transition to a liquid-crystal phase with short-range
translational order, the second does the transition to isotropic liquid.
Disorder  pins the Skyrmion lattice and also causes the crystalline order to
have a finite correlation length. For such a system provided the skyrmion
positions  fixed at all temperatures, the long-wave-length physics would be
described by an antiferromagnetic $XY$ model with expectable BKT transition and
gapless $XY$ spin-wave mode.

As regards the superfluid properties the skyrmionic liquid reveals
unconventional behavior with two critical temperatures $T_{BS}\leq t$ and $T_c
\leq J $, $T_{BS}$ being the temperature of the ordering of the modulus and
$T_c < T_{BS}$ that of the phase of order parameter $\Psi$.

The low temperature physics in Skyrmion crystals is  governed by an
interplay of two BKT transitions,  for the U(1) phase  and positional degrees of
freedom, respectively. \cite{Timm}
Dislocations in most Skyrmion lattice types lead to a mismatch in the U(1)
degree of freedom, which makes the dislocations bind fractional vortices and
lead to a coupling of translational and phase excitations. Both BKT temperatures
either coincide (square lattice) or the melting one is higher (triangular
lattice).\cite{Timm}

 Quantum fluctuations can substantially affect these
results. Quantum melting can destroy U(1) order at sufficiently
low densities where the Josephson coupling becomes exponentially small. Similar
situation is expected to take place in the vicinity of
structural transitions in Skyrmion crystal. With increasing the skyrmion density
the quantum effects  result in a significant lowering of the melting temperature
as compared with classical square-root dependence.
The resulting melting temperature can reveal  an oscilating behavior as a
function of particle density with zeros at the critical (magic) densities
associated with structural phase transitions.

In terms of our model, the positional order corresponds to an incommensurate
charge density wave, while the U(1) order does to a superconductivity. In other
words, we arrive at a subtle interplay between two orders. The superconducting
state evolves from a charge order with $T_C \leq T_m$, where $T_m$ is the
temperature of a melting transition which could be termed as a temperature of
the opening of the insulating gap (pseudo-gap!?).

The normal modes of a dilute skyrmion  system (multi-center skyrmion)
include the pseudo-spin waves
propagating in-between the skyrmions, the positional fluctuations, or phonon
modes, of the skyrmions which are gapless in  a pure system, but gapped  when
the lattice is pinned, and, finally,  fluctuations in the skyrmionic in-plane
orientation and  size.
The latter two types of fluctuation are intimitely
connected, since the $z$-component of
   spin and  orientation are conjugate coordinates because of  commutation
relations of quantum
   angular momentum operators. So, rotating a skyrmion changes its size.
    The orientational
   fluctuations of the multi-skyrmion system are governed by the gapless
   $XY$ model.\cite{Green} The relevant model description is most familiar as
an effective   theory of the Josephson junction array. An important feature of
the model is that it displays a quantum-critical point.

The low-energy collective excitations of skyrmion
liquid includes an usual longitudinal acoustic phonon branch.
The liquid crystal phases differ from the isotropic liquid in that they have
massive topological excitations, {\it i.e.}, the disclinations.
One should note that the liquids do not support transverse modes, these could
survive in a liquid state only as overdamped modes.  So that it is reasonable to
assume that solidification of the skyrmion lattice would be accompanied by a
stabilization of transverse modes with its sharpening below melting transition.
In other words an instability of transverse phonon modes signals the
onset of melting.

A generic property of the positionally ordered skyrmion configuration is the
sliding mode which is usually pinned by the disorder. The depinning of sliding
mode(s) can be detected in a low-frequency and low-temperature optical response.

In conclusion, the boson/hole doping of the
hard-core boson system  away from half-filling  is assumed to be a driving force
for a nucleation of  a  multi-center skyrmion-like self-organized
collective mode that resembles  a system of CO
bubble domains with a Bose superfluid and extra bosons both confined in domain
walls. Such a  {\it topological} CO+BS {\it phase separation}, rather than an
uniform mixed CO+BS
supersolid phase, is believed to describe the evolution of hc-BH model away from
half-filling. Starting from the classical model we predict the properties
of the respective quantum system. In frames of our scenario we may anticipate
for the hc-BH model the emergence
 of an inhomogeneous BS condensate for superhigh temperatures $T_{TPS}\leq t$,
  and 3D superconductivity for rather high temperatures $T_{c}\leq J < t$. The
system is
believed to reveal many properties typical for granular superconductors, CDW
materials,  Wigner crystals, and multi-skyrmion system akin in a quantum Hall
ferromagnetic state of a 2D electron gas.
  Topological inhomogeneity is believed  to be a generic property of 2D
  hard-core boson systems away from half-filling.

Despite all shortcomings, MFA and continuous approximation  are expected to
provide a physically clear
semiquantitative picture  of rather complex transformations taking
place in bare CO system with doping, and can be instructive as a starting point
to analyze possible scenarios. First of all, the MFA analysis allow us to
consider the  antiphase domain wall in CO phase to be a very
efficient ring-shaped potential well for the localization of a single extra
boson (hole) thus forming a novel type of a topological defect with a
single-charged domain wall. Such a defect can be addressed as a charged
skyrmion-like quasiparticle which energy can be approximated by its classical
value for CO bubble domain.
It is of great importance to note that domain wall simultaneously
represents a ring-shaped reservoir for Bose superfluid.

 Unfortunately, we have no experience to deal with multi-center
skyrmions as regards its structure, energetics, and stability.  It should be
noted that such a texture with strongly polarizable centers is believed to
provide an effective screening of
long-range boson-boson repulsion thus  resulting in an additional
self-stabilization.
Nucleation of topological phase is likely to proceed in the way typical for the
first order phase transitions.
The present paper establishes only the framework for analyzing the  subtleties
of the phase separation in a lattice hc-BH model away from half-filling. Much
work remains to be done both in a macroscopic and microscopic aproaches.

 We acknowledge the support by  SMWK Grant, INTAS Grant No. 01-0654, CRDF Grant
No. REC-005,
RME Grant No. E 02-3.4-392 and No. UR.01.01.042, RFBR Grant No. 01-02-96404.
One of us (A.S.M.) has benefited from discussions with C. Timm,  S.-L.
Drechsler, and T. Mishonov.


\begin{thebibliography}{99}


\bibitem{Penrose}
O. Penrose and L. Onsager, Phys. Rev. {\bf 104}, 576 (1956).



 \bibitem{Batrouni}
 G. G. Batrouni and R. T. Scalettar, Phys. Rev. Lett. {\bf 84}, 1599
 (2000).

\bibitem{Hebert}
 F. H\'{e}bert, G.G. Batrouni, R.T. Scalettar {\it et. al.},
Phys. Rev. B {\bf 65}, 014513 (2001)

\bibitem{Schmid}
Guido Schmid, Synge Todo, Matthias Troyer, and Ansgar Dorneich,
 Phys. Rev. Lett. {\bf 88}, 167208 (2002).

 \bibitem{MFA}
 R.T. Scalettar, G.G. Batrouni, A.P. Kampf, and G.T. Zimanyi,
 Phys. Rev. B {\bf 51}, 8467 (1995).

\bibitem{Cagan}
M.Yu. Kagan, K.I. Kugel, and D.I. Khomskii, JETP, {\bf 93}, 415 (2001).


\bibitem{Pich}
Christian Pich and Erwin Frey, Phys. Rev. B {\bf 57}, 13712 (1998).

\bibitem{Bernardet}
 K. Bernardet, G.G. Batrouni, J.-L. Meunier {\it et. al.},
Phys. Rev. B {\bf 65}, 104519 (2002).

\bibitem{AFM-domain}
 The  antiferromagnetic domain texture appears as
a result of the minimization of elastic and magnetoelastic energies.

\bibitem{Belavin} A.A. Belavin, A.M. Polyakov, JETP Lett. {\bf 22}, 245 (1975).

\bibitem{Waldner}
F. Waldner,  J. Magn. Mag. Matter {\bf 54-57}, 837 (1986); Phys. Rev. Lett. {\bf
65}, 1519 (1990).

\bibitem{Kochelaev}
S.I. Belov, B.I. Kochelaev, Sol. St. Commun. {\bf 103}, 249 (1997).

\bibitem{Carsten}
Carsten Timm and K. H. Bennemann, Phys. Rev. Lett. {\bf 84}, 4994 (2000).


\bibitem{Kamppeter}
T. Kamppeter, S.A. Leonel, F.G. Mertens, M.E. Gouvêa, A.S.T. Pires, and A.S.
Kovalev,  Eur. Phys. J. B {\bf 21}, 93 (2001).

\bibitem{Sheka}
Denis D. Sheka, Boris A. Ivanov, and G. Mertens,
Phys. Rev. B {\bf 64}, 024432 (2001).

\bibitem{Istomin}
R.A. Istomin, A.S. Moskvin, JETP Lett. {\bf 61}, 898 (2000).


\bibitem{AFM}
 A.S. Ovchinnikov, I.G. Bostrem, A.S. Moskvin, Phys. Rev. B {\bf 66},
 134304 (2002).

\bibitem{Abanov}
Ar. Abanov, V.L. Pokrovsky, Phys. Rev. B {\bf 58}, R8889 (1998).


\bibitem{Green}
A.G. Green, Phys. Rev. B {\bf 61}, R16299 (2000).

\bibitem{Timm}
Carsten Timm, S.M. Girvin, H.A. Fertig, Phys. Rev. B {\bf 58},  10634 (1998).

\bibitem{Kivelson}
S.A. Kivelson, B.Z. Spivak, Phys. Rev. B {\bf 45}, 10490 (1992).

\bibitem{Rao}
Madan Rao, Surajit Sengupta, and R. Shankar, Phys. Rev. Lett. {\bf 79}, 3998
(1997).

\end{thebibliography}
\end{document}